\documentclass[a4paper,fleqn,usenatbib]{mnras}
\usepackage[T1]{fontenc}
\usepackage{aecompl}
\usepackage{graphicx}
\usepackage{epstopdf}
\usepackage{amsmath}
\usepackage{amssymb}	
\usepackage{enumerate}

\title[Quantifying the Colour-Dependent Stochasticity of Large-Scale Structure]{Quantifying the Colour-Dependent Stochasticity of Large-Scale Structure}
\author[A. Patej \& D. Eisenstein]{
Anna Patej$^{1}$ and 
Daniel Eisenstein$^{2}$ \\
$^{1}$Department of Physics, Harvard University, 17 Oxford St., Cambridge, MA 02138, USA\\
$^{2}$Harvard-Smithsonian Center for Astrophysics, 60 Garden St., Cambridge, MA 02138, USA
}

\begin{document}

\pagerange{\pageref{firstpage}--\pageref{lastpage}}\pubyear{2015}
\maketitle
\label{firstpage}

\begin{abstract}
We address the question of whether massive red and blue galaxies trace the same large-scale structure at $z\sim0.6$ using the CMASS sample of galaxies from Data Release 12 of the Sloan Digital Sky Survey III. After splitting the catalog into subsamples of red and blue galaxies using a simple colour cut, we measure the clustering of both subsamples and construct the correlation coefficient, $r$, using two statistics. The correlation coefficient quantifies the stochasticity between the two subsamples, which we examine over intermediate scales ($20 \lesssim R \lesssim 100\;h^{-1}\mathrm{Mpc}$). We find that on these intermediate scales, the correlation coefficient is consistent with 1; in particular, we find $r > 0.95$ taking into account both statistics and $r > 0.974$ using the favored statistic.
\end{abstract}

\section{Introduction}\label{s:intro}

Large-scale galaxy redshift surveys such as the Sloan Digital Sky Survey \citep[SDSS-III DR12;][]{sdssdr12} use the distribution of galaxies in the universe to constrain cosmological parameters in a manner complementary to other cosmological probes, including measurements of the cosmic microwave background \citep[e.g.,][]{planck_cosmo} and supernovae \citep[e.g.,][]{suzuki12}. Within the scope of galaxy redshift surveys, major projects like SDSS-III's Baryon Oscillation Spectroscopic Survey \citep[BOSS;][]{dawson13} as well as future experiments like the Dark Energy Spectroscopic Instrument (DESI) \citep{levi13} target specific types of galaxies: luminous red galaxies (LRGs), and LRGs and emission line galaxies (ELGs), respectively. Accordingly, a key source of systematic uncertainty in these surveys is the knowledge of the extent to which the subset of galaxies observed probes the large-scale structure of the universe. 

Local bias argues for a minimal level of stochasticity on large scales \citep{coles93,scherrer98} and investigations of halo clustering suggest that halos of different masses do roughly trace the same large-scale structure \citep[e.g,][]{seljak04,hamaus10}. Accordingly, we may expect galaxies to be roughly non-stochastic \citep[e.g.,][]{tegmark99}. An informative test of these predictions can be obtained by measuring whether red and blue galaxies trace the same large-scale structure. The mathematical framework of this question can be developed straightforwardly using the concept of a fractional overdensity field,
\begin{align}
\delta(\mathbf{x})=\frac{\rho(\mathbf{x})}{\bar{\rho}}-1.
\end{align}
In the simplest scenario, we can relate the distribution of red and blue galaxies to this underlying matter distribution via linear, deterministic bias parameters as:
\begin{align}\label{e:lindeltas}
\delta_b(\mathbf{x}) = b_b\delta(\mathbf{x}),\;\;\delta_r(\mathbf{x}) = b_r\delta(\mathbf{x}),
\end{align}
from which it is possible to compute the correlation function as $\xi(\mathbf{R}) = \langle\delta(\mathbf{x})\delta(\mathbf{x}+\mathbf{R})\rangle$. Combined with Equation~(\ref{e:lindeltas}), this yields the relations:
\begin{align}
\xi_{bb}(\mathbf{R}) &= b_b^2\xi(\mathbf{R}),\\
\xi_{rr}(\mathbf{R}) &= b_r^2\xi(\mathbf{R}),\\
\xi_{br}(\mathbf{R}) &= b_{b}b_r\xi(\mathbf{R}).
\end{align}
Taking the square root of the ratios of the autocorrelations yields estimates of the relative bias $b_\mathrm{rel}$ between red and blue galaxies \citep[e.g.,][]{croton07,coil08,guo13,skibba14}. Constructing the correlation coefficient, $r_{\xi}$, we find:
\begin{align}
r_{\xi} \equiv \frac{\xi_{br}}{\sqrt{\xi_{bb}\xi_{rr}}} =1.
\end{align}

However, the formalism of Equation~(\ref{e:lindeltas}) has several obvious failures, one of which is that it permits values of $\delta_b, \delta_r < -1$ if $b_b,b_r > 1$, and so must be superseded by a more realistic model. To this end, we follow \cite{dekel99} in defining, for $g=b,r$,
\begin{align}\label{e:dekeleq}
\epsilon_g(\mathbf{x}) \equiv \delta_g(\mathbf{x})-b_g\delta(\mathbf{x}),
\end{align}
a random bias field that introduces stochasticity into the relations between the two galaxy samples. In this case, if we calculate the correlation coefficient, we see that $r_{\xi} \ne1$.

The key to discerning the presence of stochasticity, then, is the measurement of the correlation coefficient. We use pair counting to calculate both the traditional correlation function statistics as well as the more recent $\omega$ statistic of \cite{xu10} using the BOSS CMASS sample of galaxies from SDSS-III DR12; we then calculate a correlation coefficient for each statistic. Our focus is on intermediate scales, roughly $20\lesssim r\lesssim100\;h^{-1}\mathrm{Mpc}$, which can be compared to the results of previous analyses at smaller scales using similar statistical methods \citep[e.g.,][]{zehavi05}. Our approach also provides an alternative to analyses of colour-dependent clustering using the complementary counts-in-cells method \citep[e.g.,][]{wild05,swanson08}. Throughout we assume a flat $\Lambda$CDM cosmology with $\Omega_m = 0.274$, which is consistent with \citet{anderson12}.

\section{Data}\label{s:data}
The SDSS-III DR12 contains spectra of over 1.3 million galaxies and $ugriz$ imaging of 14555 sq. degrees of sky obtained using a 2.5 m telescope at Apache Point Observatory \citep{york00,gunn06,sdssdr12}. A series of publications outlines the technical details of SDSS instrumentation \citep{fukugita96,smith02,doi10,smee13} and the data processing pipelines \citep{lupton01,pier03,padmanabhan08,bolton12,weaver15}. We perform our analysis on the CMASS sample of BOSS galaxies, which is defined via colour and magnitude cuts as in \citet{eisenstein11}, \citet{dawson13}, and \citet{reid15}. We further narrow our attention to the redshift range $0.55<z<0.65$. To divide the sample into red and blue galaxies, we use the criterion of \cite{masters11}, which selects red galaxies using the simple colour cut $g-i > 2.35$. These cuts yield 232,759 red galaxies and 61,301 blue galaxies. In addition to the data, we select two subsamples of random galaxies to correspond to the red and blue galaxies such that each set has roughly 50 times the number of galaxies as the data subsamples. 

\begin{figure*}
\begin{center}
\includegraphics[scale=0.51,trim={0.45cm 0.4cm 0.2cm 0.2cm},clip]{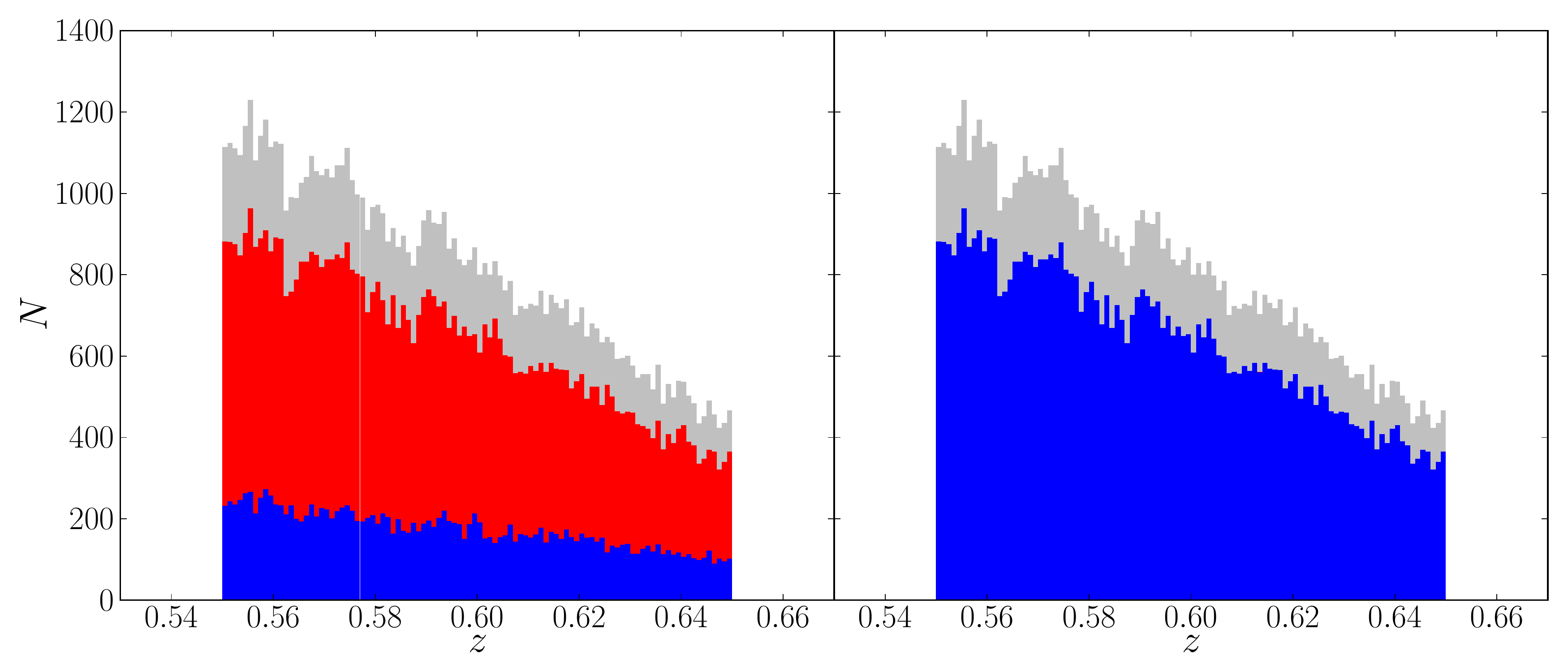}
\end{center}
\caption{A histogram of the distribution of galaxies in the DR12 CMASS South sample over $0.55<z<0.65$. The lefthand panel shows the original division into red and blue galaxies following the simple colour cut described in Section \ref{s:data}; in the righthand panel, we show the distribution after weighting the blue sample to match the red galaxies.}\label{f:zdist}
\end{figure*}

Since this colour cut yields roughly four times more red galaxies than blue, we use the redshift distribution of galaxies, shown in Fig.~\ref{f:zdist}, to generate weights $w_{\mathrm{colour}}$ for the blue galaxies that match their distribution to the red galaxies, which we will use in addition to the standard data weights. The final weighting that we use is then:
\begin{align}
w_{\mathrm{total}} = w_{\mathrm{colour}}w_{\mathrm{sys}}(w_{zf}+w_{cp}-1),
\end{align}
where $w_{\mathrm{colour}}=1$ for red galaxies and $w_{\mathrm{colour}}>1$ for blue galaxies, $w_{\mathrm{sys}}$ is a systematic weight that accounts for observing conditions, and $w_{zf}$ and $w_{cp}$ account for redshift failures and close pairs, respectively \citep[c.f.][]{anderson12,ross14}. We additionally match the redshift distribution of the randoms to that of the data.

It is worth noting that the simple colour cut of \cite{masters11} is not the only choice for dividing the galaxy sample into red and blue subsets; other works have employed different cuts, such as the luminosity dependent colour cut of \citet{zehavi11} and \citet{guo13}. However, for the purposes of this work, the exact choice of the colour cut is not significant, since we make the randoms trace the galaxy redshift distributions. In essence, all we require is a simple criterion by which to divide the sample into two categories, which we compare via two statistics. 

\section{Procedure}\label{s:method}
\subsection{Statistics}
One of the standard methods of calculating the galaxy correlation function is the Landy-Szalay estimator~\citep{landy93}:
\begin{align}\label{e:ls}
\xi(r) = \frac{DD(r)-2DR(r)+RR(r)}{RR(r)},
\end{align}
where the terms $DD$, $DR$, and $RR$ denote normalised data-data, data-random, and random-random pair counts. In analyzing red and blue galaxy clustering, we may thus define the auto-correlations of blue galaxies and red galaxies as:
\begin{align}
\xi_{bb}(r) = \frac{D_{b}D_{b}(r)-2D_{b}R_{b}(r)+R_{b}R_{b}(r)}{R_{b}R_{b}(r)},\\
\xi_{rr}(r) = \frac{D_{r}D_{r}(r)-2D_{r}R_{r}(r)+R_{r}R_{r}(r)}{R_{r}R_{r}(r)},
\end{align}
and the cross-correlation (blue-red) as:
\begin{align}
\xi_{br}(r) = \frac{D_{b}D_{r}(r)-D_{b}R_{r}(r)-D_{r}R_{b}(r)+R_{b}R_{r}(r)}{R_{b}R_{r}(r)}.
\end{align}
From these functions we can then compute the cross-correlation coefficient,
\begin{align}
r_{\xi} = \frac{\xi_{br}}{\sqrt{\xi_{rr}\xi_{bb}}},
\end{align}
which provides a measure of the stochasticity.

However, while correlation functions provide a useful measurement of the stochasticity parameter, it is possible to extend this analysis by using a statistic that is both more computationally efficient and less susceptible to poorly constrained or measured fluctuations from both small and large scales. Such a statistic is provided by \citet{xu10}, who define an $\omega_{\ell}$ statistic whose monopole term is:
\begin{align}\label{e:omega0}
\omega_{0}(r_s) = 4\pi\int W(r,r_s)\xi(r) r^2dr,
\end{align}
where $W$ is a smooth, compensated filter, chosen by \citet{xu10} \citep[see also][]{padmanabhan07} to be:
\begin{align}
W(x) = (2x)^2(1-x)^2\left(\frac{1}{2}-x\right)\frac{1}{r_s^3},
\end{align}
where
\begin{align}
x = \left(\frac{r}{r_s}\right)^3.
\end{align}
The smoothness of the filter function $W$ means that $\omega_0$ is largely insensitive to small-scale power, while the fact that $W$ integrates to zero removes the sensitivity to large-scale power and reduces the risk of including large-scale systematic errors.

Now, following \citet{xu10}, if we define a simple pair-count estimator of the correlation function as:
\begin{align}\label{e:simplexi}
\xi(r) = \frac{DD(r)}{RR(r)}-1, 
\end{align}
then by substituting into Equation~(\ref{e:omega0}), we obtain:
\begin{align}\label{e:o1}
\omega_{0}(r_s) = 4\pi\int W(r,r_s)\frac{DD(r)}{RR(r)}r^2dr,
\end{align}
with the integral of the $-1$ term in Equation~(\ref{e:simplexi}) vanishing since we have selected a compensated filter. The DD term has a simple mathematical interpretation: $DD$ is simply a weighted count over pairs at a given separation. That is, DD can be written as a sum of Dirac delta functions:
\begin{align}
DD(r)=\sum_j\eta_j\delta_D(r-r_j),
\end{align}
where $\eta_j$ is the product of the $w_\mathrm{tot}$ of the two galaxies in the pair. The $RR$ term, on the other hand, is a description of the survey region, and may be written as:
\begin{align}\label{e:RRphi}
RR(r) = N_D^2\Phi(r)\frac{4\pi r^2 dr}{V}.
\end{align}
Here, $N_D$ is the weighted number of (data) galaxies, and $V$ is the total survey volume. $\Phi(r)$ is a function that takes into account the survey boundaries; that is, it measures the fraction of pairs that are lost because the central galaxy around which the pair counts are being computed is close to the boundaries. $\Phi(r)$ can be approximated by counting up $RR$ pairs and then fitting a polynomial to the formula defined in Equation~(\ref{e:RRphi}). With $\Phi$ thus calculated, rearranging this equation yields:
\begin{align}\label{e:rearrange}
\frac{V}{N_D^2}\frac{1}{\Phi(r)}=\frac{4\pi r^2dr}{RR(r)}.
\end{align}

Returning to Equation~(\ref{e:o1}), we find:
\begin{align}
\omega_{0}(r_s) &= \int W(r,r_s)\sum_j\eta_j\delta_D(r-r_j)\frac{4\pi r^2}{RR(r)}dr.
\end{align}
The $\delta_D$ function then collapses the integral so that, upon substituting Equation~(\ref{e:rearrange}) and absorbing $\eta_j$ into $W$, we obtain the following summation:
\begin{align}
\omega_{0}(r_s) = \frac{V}{N_D^2}\sum_{j}\frac{W(r_j,r_s)}{\Phi(r_j)}.
\end{align}

Accordingly, we see that the $\omega_0$ statistic reduces to a discrete sum over pairs. So defined, $\omega_0$ reveals several additional advantages over the traditional $\xi$ statistic \citep{padmanabhan07,xu10}. First, when computing $\xi$, we initially bin the data to compute the pair counts at a discrete set of points. When computing $\omega_0$, on the other hand, we sum over each pair individually, which eliminates binning concerns. Additionally, doing a count over just data pairs is more computationally efficient than performing three sets of counts ---  $DD$, $DR$, and $RR$ --- particularly since the number of randoms is typically chosen to be about 50 times the number of data points for computing $\xi$; while $RR$ still needs to be computed in order to obtain $\Phi$, one can use far fewer points to get an estimate of the geometry of the survey. 

Now, we can define $\omega$ analogues to all the correlation function statistics. The correlation functions themselves are translated into:
\begin{align}
\omega_{bb}(r_s) &= \frac{V}{N_b^2}\sum_{j\in{D_bD_b}}\frac{W(r_j,r_s)}{\Phi(r_j)},\\
\omega_{rr}(r_s) &= \frac{V}{N_r^2}\sum_{j\in{D_rD_r}}\frac{W(r_j,r_s)}{\Phi(r_j)},\\
\omega_{br}(r_s) &= \frac{V}{N_bN_r}\sum_{j\in{D_bD_r}}\frac{W(r_j,r_s)}{\Phi(r_j)}.
\end{align}
From these, we construct an $\omega$ coefficient analogous to $r_{\xi}$ as:
\begin{align}
r_{\omega} = \frac{\omega_{br}}{\sqrt{\omega_{bb}\omega_{rr}}},
\end{align}
which will provide a measurement of the stochasticity. In order to reduce the amount of subscripts in certain contexts, we note that we will use $\omega$ and $\omega_0$ for the \citet{xu10} statistic interchangeably throughout the remainder of this work. We calculate the $\xi$ and $\omega$ statistics for three selections of random galaxies, and average the results for $r$ for our final result.

\subsection{Error Analysis}
We employ the jackknife method of error estimation to derive error bars for our measurements of $\xi$ and $\omega$. In the jackknife method, we first subdivide our survey into $N$ roughly commensurate regions, then compute the statistics $N$ times, each time leaving out one of the previously defined regions. Then, the jackknife mean of the measurements may be written:
\begin{align}
\zeta_{J,i} = \sum_{j\ne i}\frac{\zeta_j}{N-1} = \frac{N\bar{\zeta}-\zeta_i}{N-1}.
\end{align}
From this, we may derive the errors as:
\begin{align}
\sigma_{J,\bar{\zeta}}^2 = \frac{N-1}{N}\sum_{i=1}^N(\zeta_{J,i}-\bar{\zeta})^2.
\end{align}

\begin{figure}
\begin{center}
\includegraphics[scale=0.31,trim={0.2cm 0.1cm 0.2cm 0.2cm},clip]{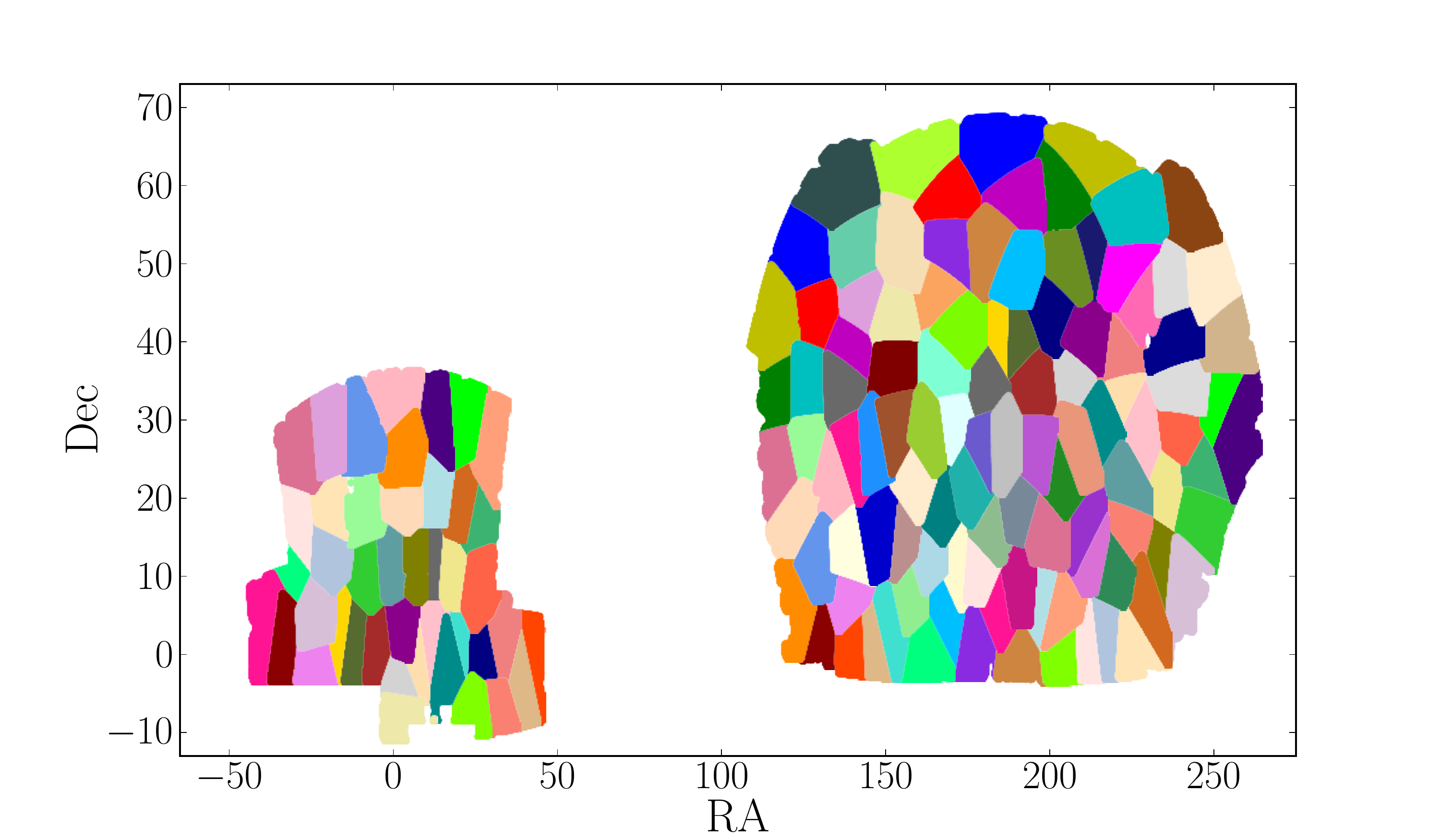}
\end{center}
\caption{Definition of 150 Voronoi regions over the entire SDSS survey area that are used in the computation of jackknife errors for one run.}\label{f:vor}\vspace{0.1in}
\end{figure}

The remaining issue is the definition of these regions. We use the method of Voronoi tessellation, in which we select a set of $N$ central points within the survey volume, and then for each of these points compute the locus of points that are closer to that central point than any other. In practice, we choose a random sample of 150 galaxies to serve as the central points. Then, to obtain regions that are similar in size (although not exactly the same; in the final analysis, the smallest and largest region differ by $\lesssim50\%$), we iterate over this selection, splitting the largest region by selecting two new central points from within that region and merging the smallest region into its neighbors by removing its central point and recomputing the regions. The result of this process is shown in Fig.~\ref{f:vor}, which shows the final set of regions for one of the three runs that contributes to our final averaged value for $r$. Each of the final Voronoi regions spans an area of roughly $14555/150\approx 100\;\mathrm{sq.\;degrees}$, or roughly 10 degrees on each side, which amounts to a transverse comoving distance of roughly $1.6\; h^{-1}\mathrm{Gpc}$ at $z=0.6$, the midpoint of our redshift range. 

\section{Results}\label{s:results}
\begin{figure*}
\includegraphics[scale=0.40,trim={0.1cm 1cm 1cm 1cm},clip]{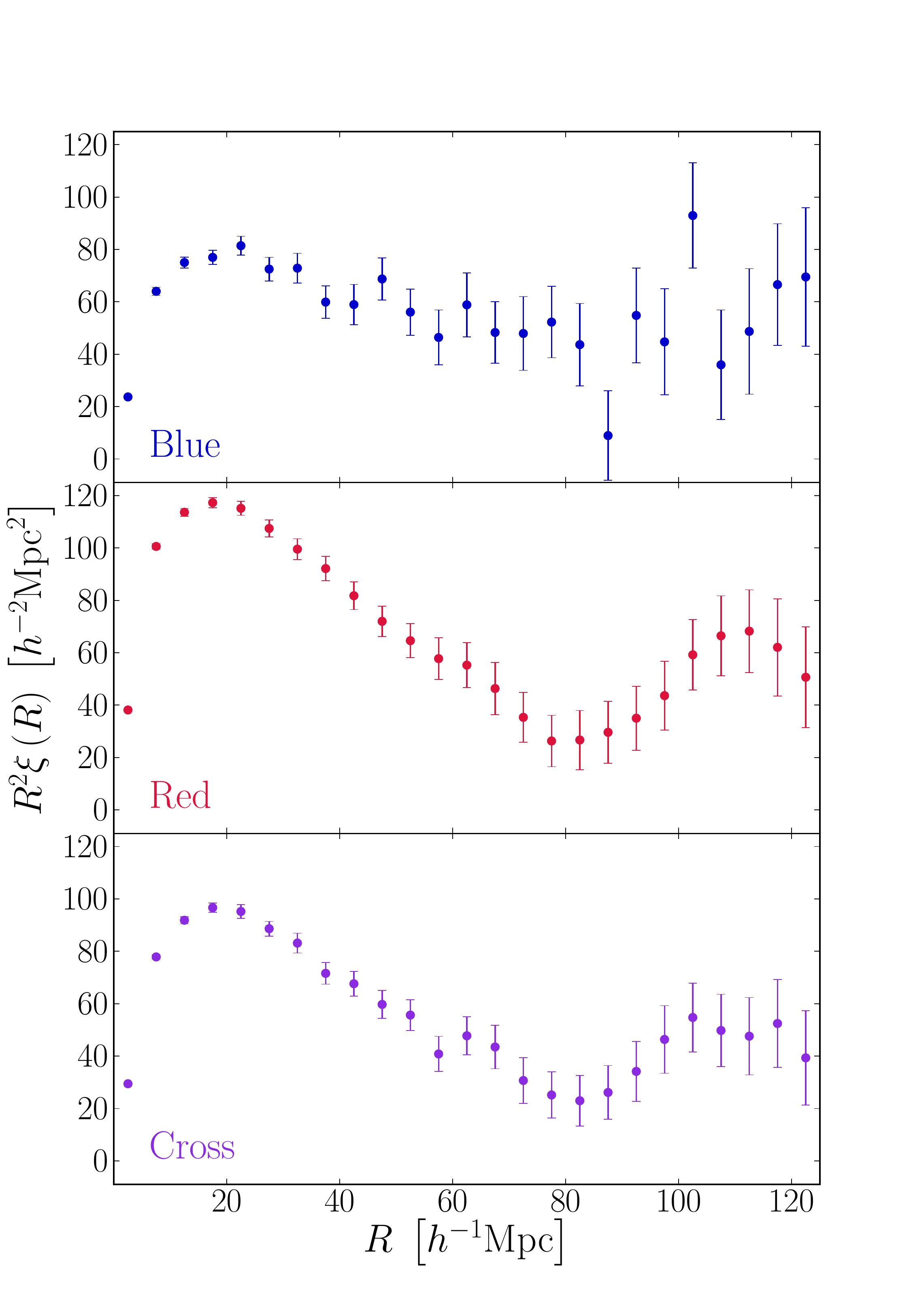}
\includegraphics[scale=0.40,trim={0.1cm 1cm 1cm 1cm},clip]{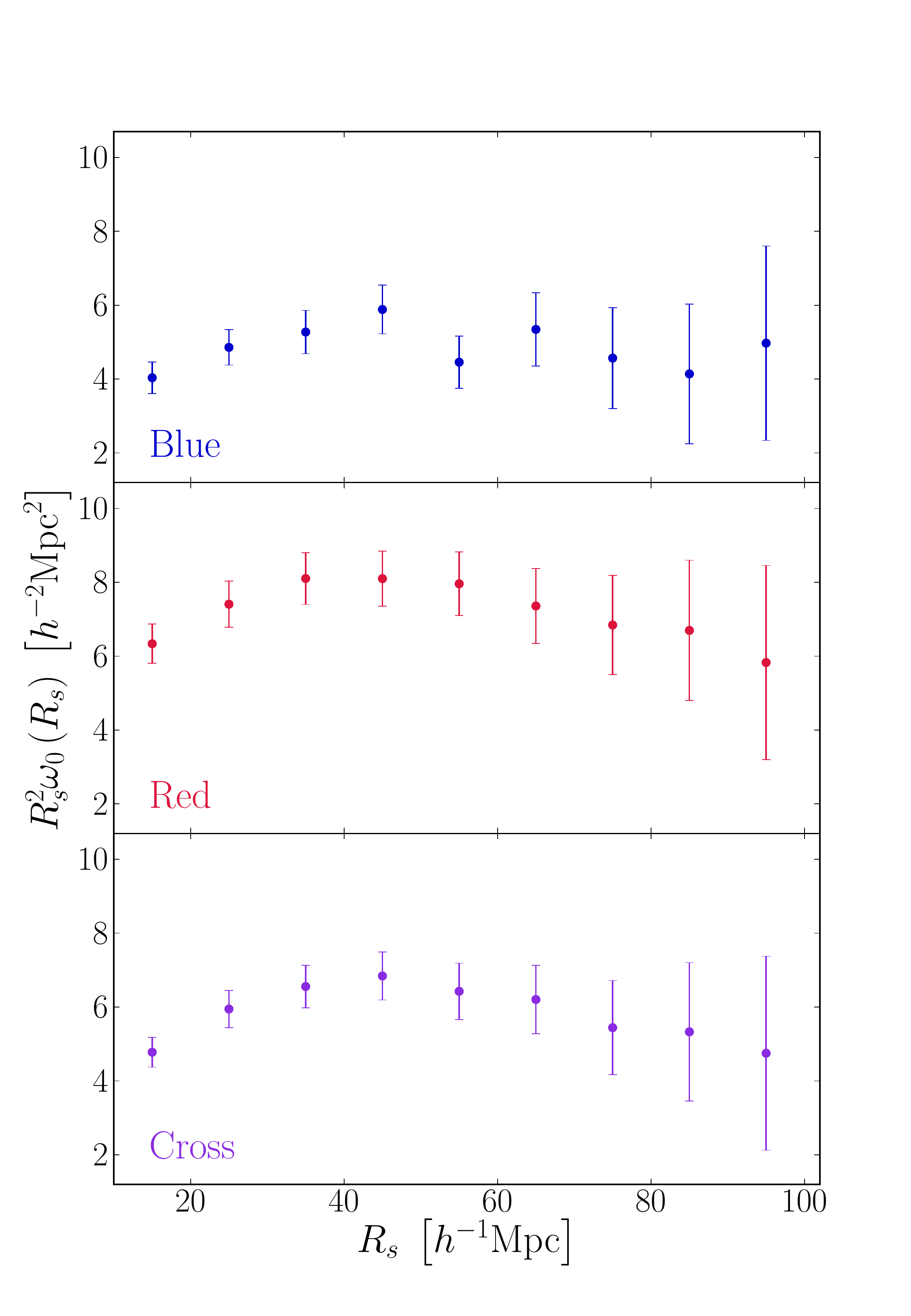}
\caption{Left: Blue-blue (top), red-red (middle), and blue-red (bottom) correlation functions, computed using a bin size of $5\;h^{-1}$ Mpc and scaled by $R^2$. Right: Blue-blue, red-red, and blue-red $\omega_0$ functions, computed using an $R_s$ spacing of $10\;h^{-1}$ Mpc and scaled by $R_s^2$.}\label{f:xu_all}
\end{figure*}

We compute both the $\xi$ and $\omega$ statistics for the CMASS sample of galaxies in three iterations, each with a different selection of randoms and of Voronoi regions. For the correlation function, we choose a binning of $\Delta R = 5\;h^{-1}\mathrm{Mpc}$, while for $\omega$, we select 9 values of $R_s$ for which to calculate the statistic, each spaced by $10\;h^{-1}\mathrm{Mpc}$. The correlation functions --- the blue and red auto-correlations, and the blue-red cross-correlation --- and the corresponding $\omega$ statistics are computed in Fig.~\ref{f:xu_all} for one of the runs.

\begin{figure*}
\begin{center}
\includegraphics[scale=0.49]{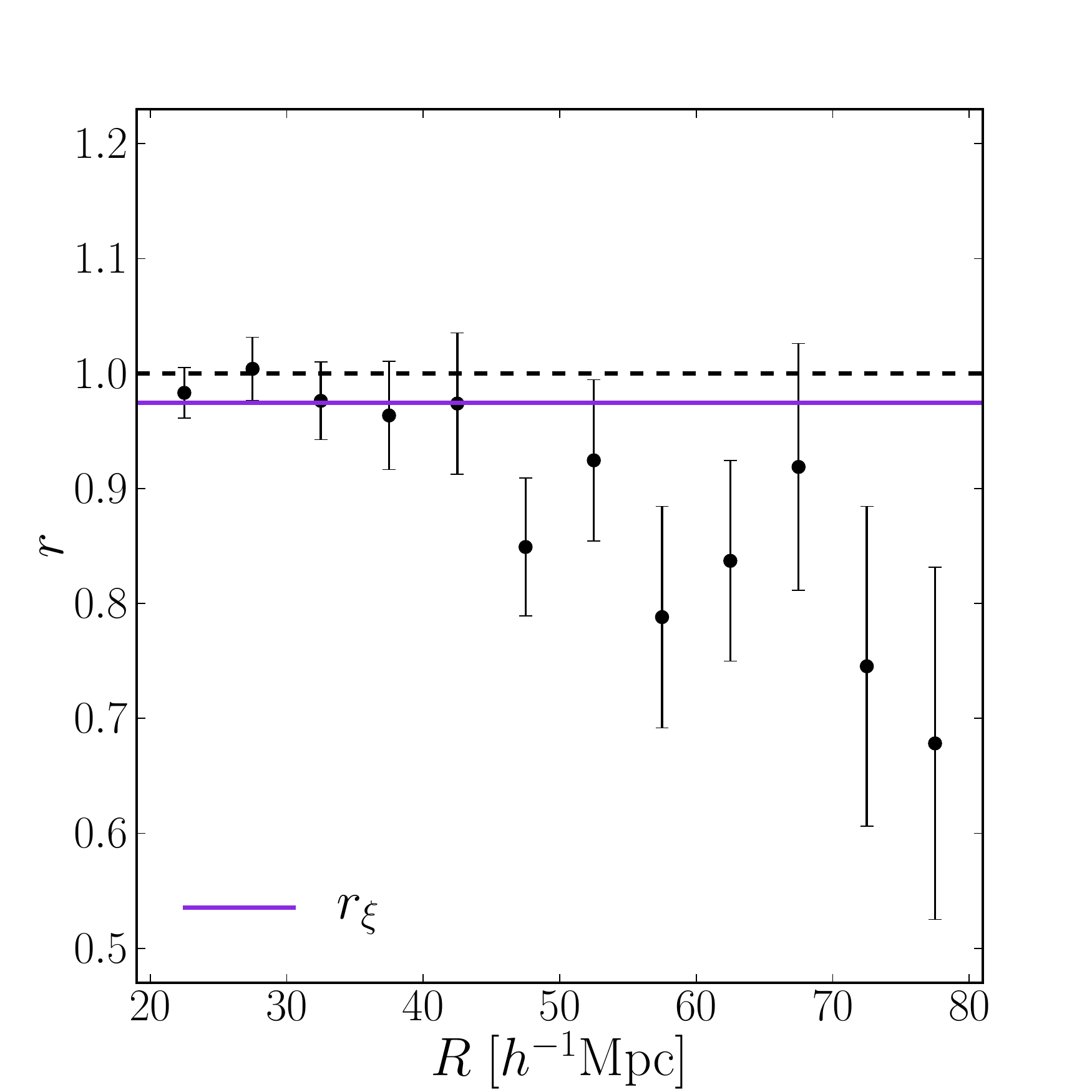}
\includegraphics[scale=0.49]{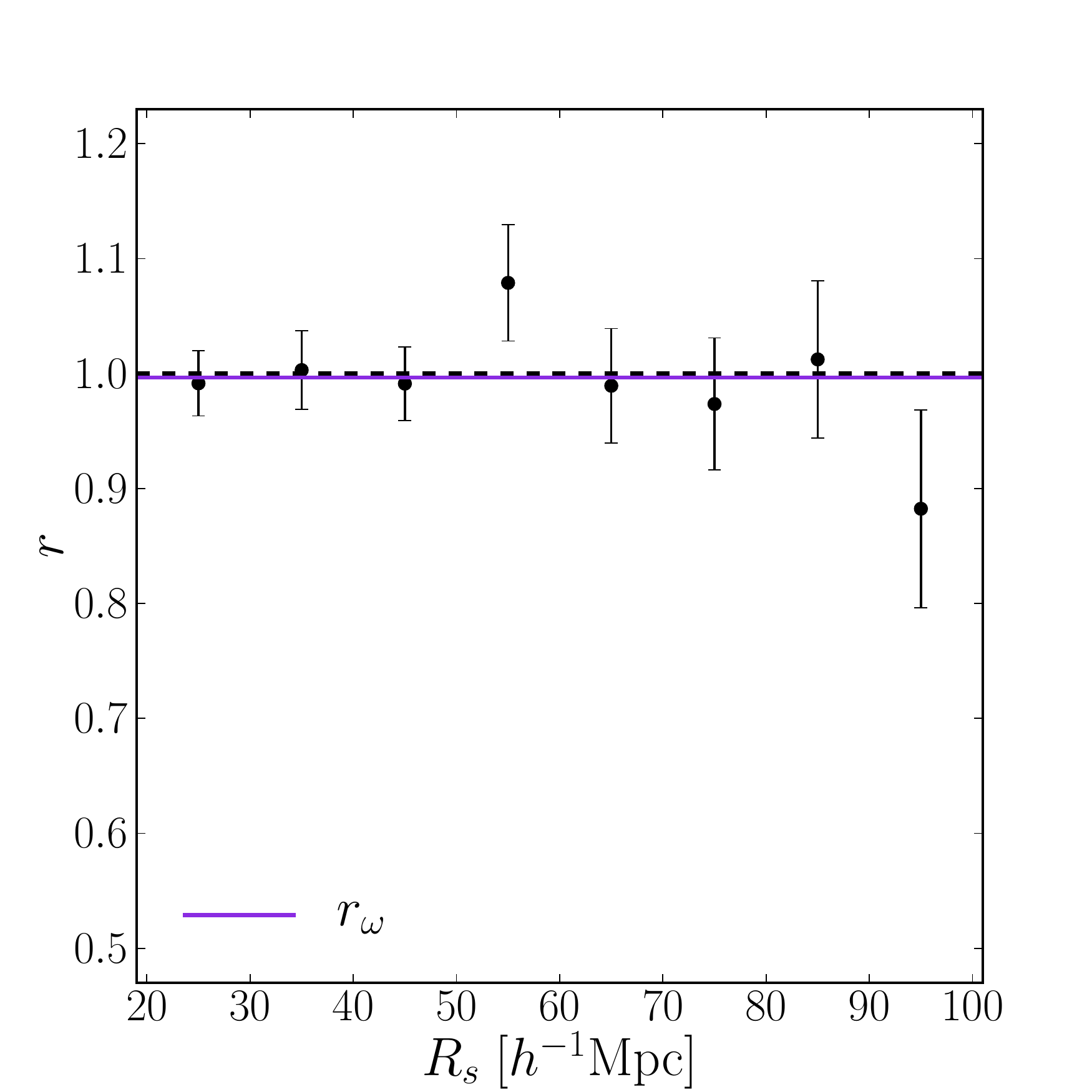}
\end{center}
\caption{The correlation coefficient, $r_{\xi}$ (left) and the analogous coefficient $r_{\omega}$ (right) given by the best fit line (solid) from one run. A dashed line is drawn at $r=1$ for comparison. Our results using the $\omega$ statistic indicate less stochasticity on intermediate scales than the noisier analysis using correlation functions. }\label{f:xi_r}
\end{figure*}

\begin{figure*}
\begin{center}
\includegraphics[scale=0.470]{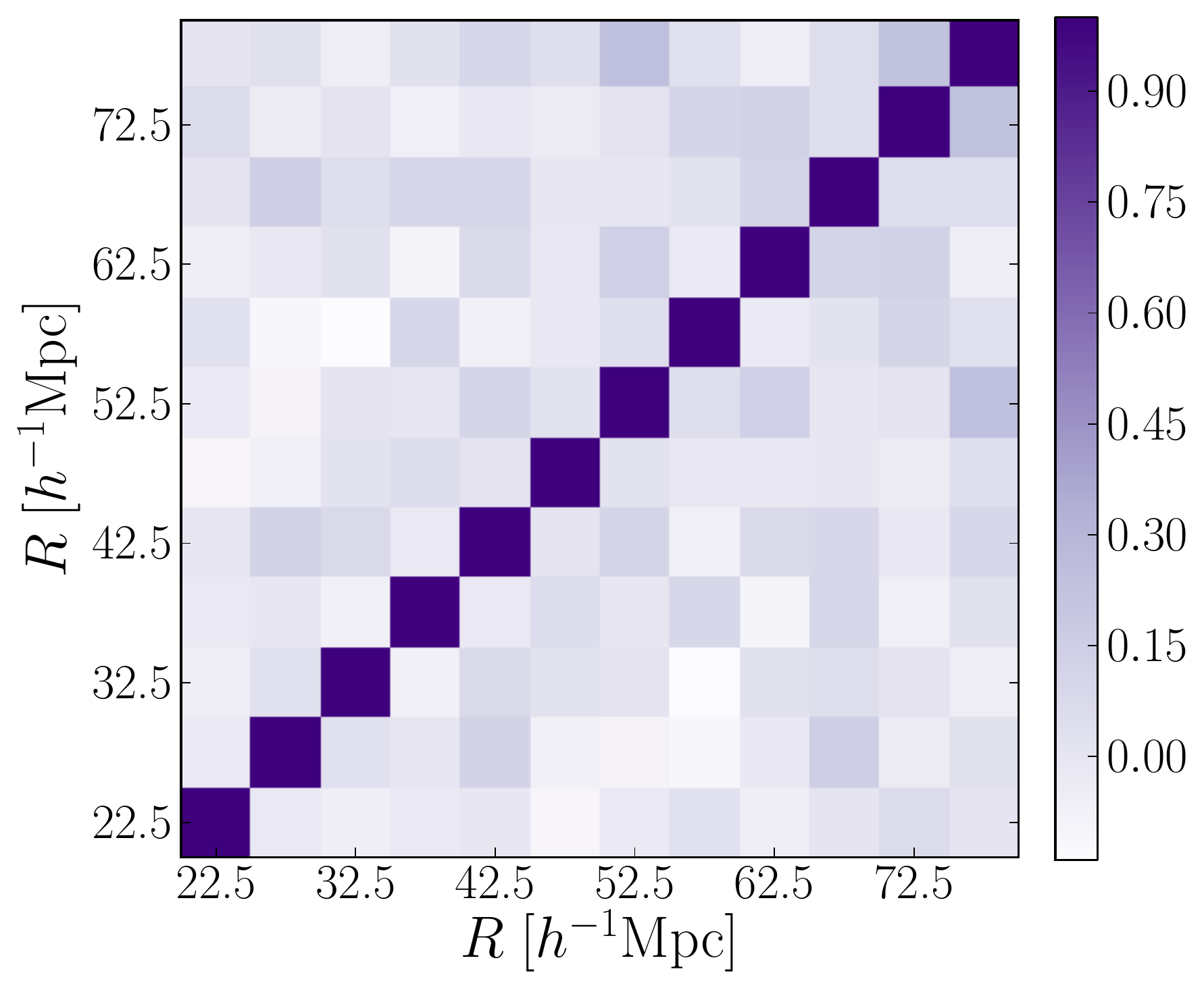}
\includegraphics[scale=0.470]{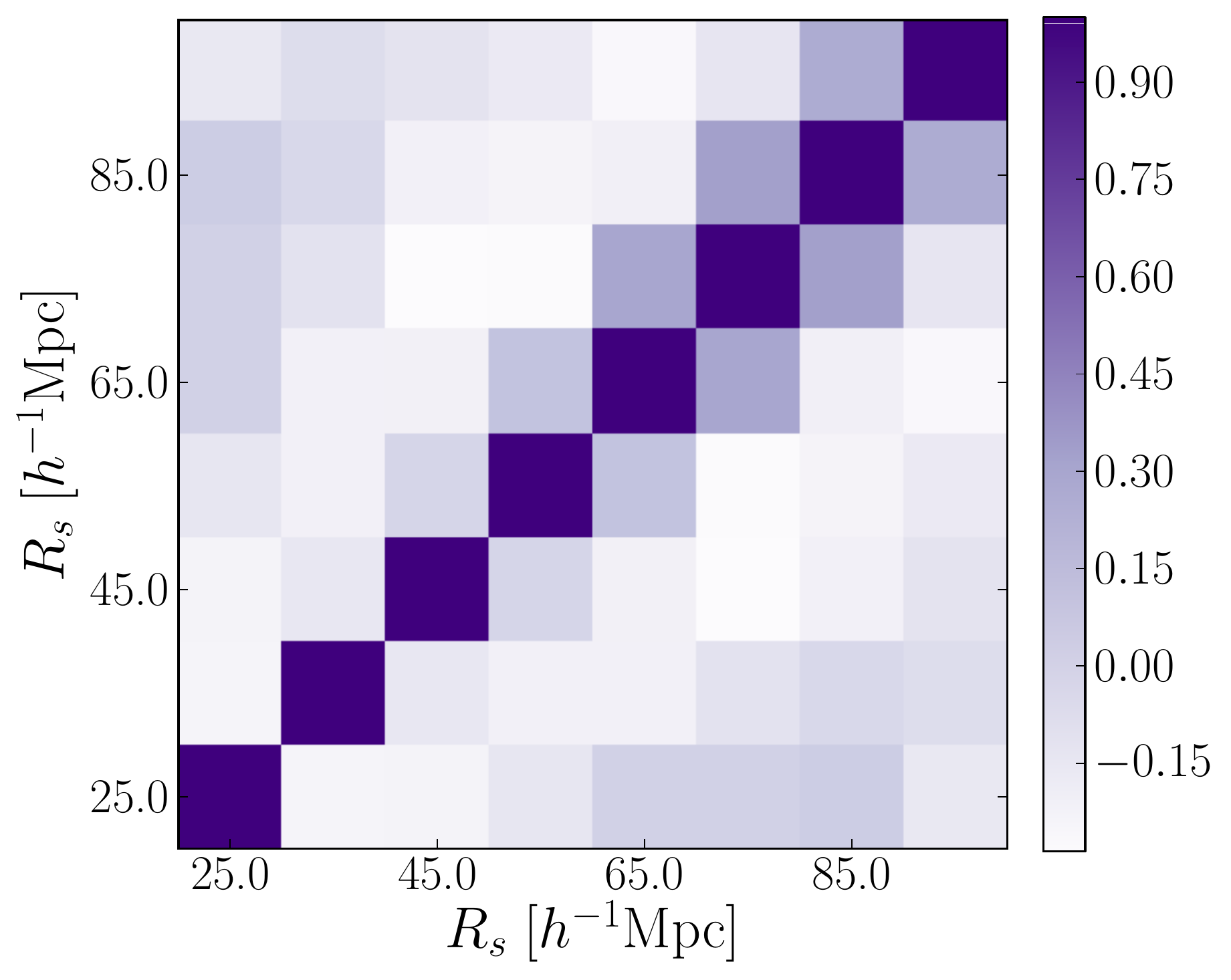}
\end{center}
\caption{The reduced covariance matrices for the correlation coefficient,  $r_{\xi}$ (left), and the analogous coefficient, $r_{\omega}$ (right), calculated for $R < 80 \;h^{-1}\mathrm{Mpc}$ and $R_s < 100 \;h^{-1}\mathrm{Mpc}$ from the run whose results are shown in Fig.~\ref{f:xi_r}. These matrices are used to calculate the $\chi^2$ for fitting $r$.}\label{f:xu_cov}
\end{figure*}

From the three correlation functions, we obtain the correlation coefficient $r_{\xi}$, shown in the left-hand panel of Fig.~\ref{f:xi_r} for one of the iterations, and fit to it a horizontal line. Although we compute the correlation functions themselves out to large scales, for the calculation of the correlation coefficient, we use the range $20 < R < 80\;h^{-1}\mathrm{Mpc}$. We also fit the analogous coefficient $r_{\omega}$ (albeit in the slightly extended range $20 < R_s < 100\;h^{-1}\mathrm{Mpc}$), the results of which can be seen in the righthand panel of Fig.~\ref{f:xi_r} for one of the runs. The reduced covariance matrices used in that iteration are furnished in Fig.~\ref{f:xu_cov}. It is worth noting that the $\xi$ covariance matrix is highly diagonal; this is likely due to shot noise in the blue subsample of galaxies. The $\omega$ covariance matrix, on the other hand, does demonstrate some covariance due to integrating $\xi$ over overlapping kernels.

We average the stochasticity parameters $r$ that we measure this way from the three iterations; the resulting values are:
\begin{align}
\langle r_{\xi} \rangle &= 0.978\pm0.014,\\
\langle r_{\omega} \rangle &= 0.996\pm0.011.
\end{align}
Although $r_{\xi}$ is lower than $r_{\omega}$, the results of both suggest low stochasticity on intermediate scales. The difference in these values is possibly influenced by the substantial scatter for $R\gtrsim 45\;h^{-1}\mathrm{Mpc}$ in $r_{\xi}$, to which the $\omega$ statistic is less susceptible. 

\begin{figure*}
\begin{center}
\includegraphics[scale=0.49,trim={0cm 0cm 0cm 1cm},clip]{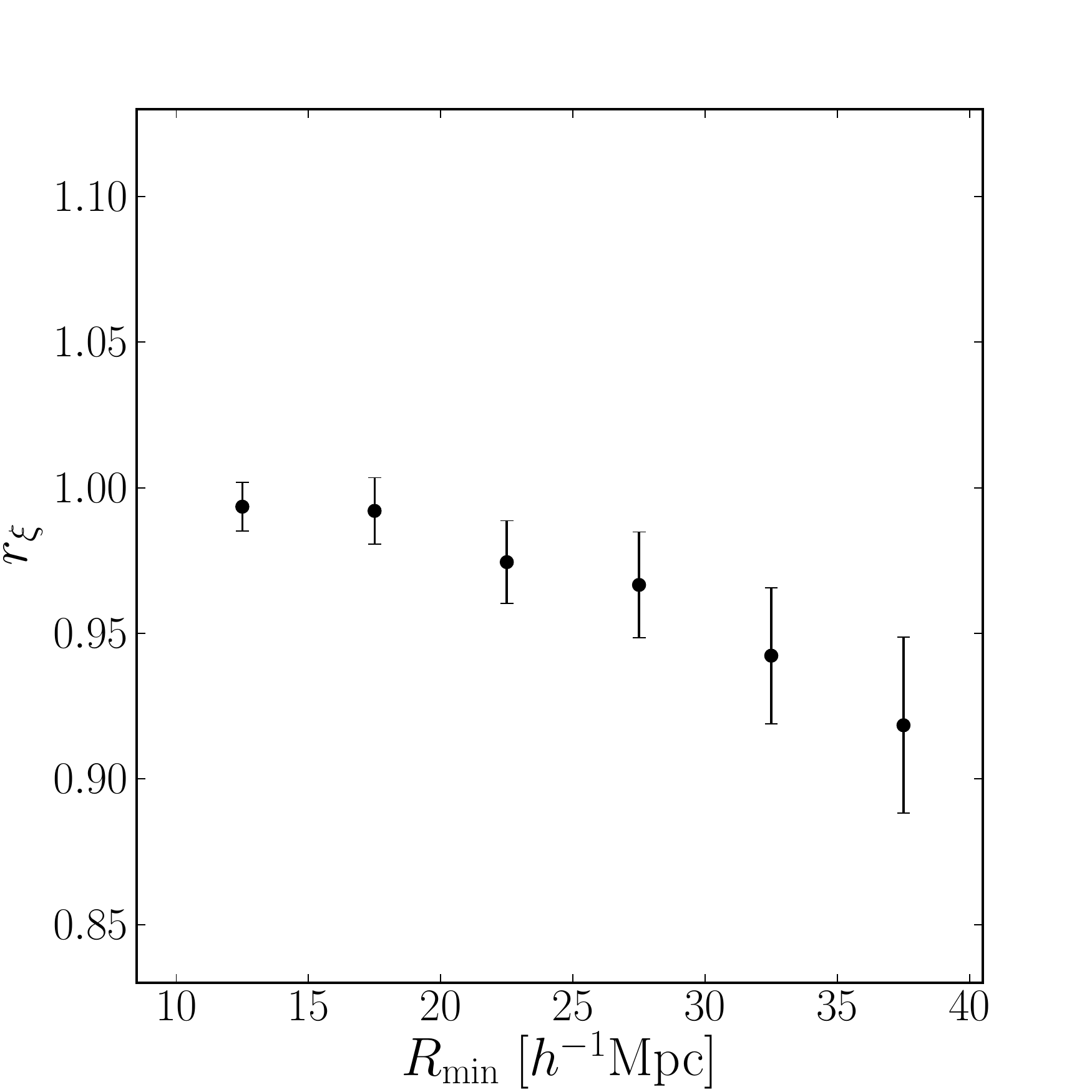}
\includegraphics[scale=0.49,trim={0cm 0cm 0cm 1cm},clip]{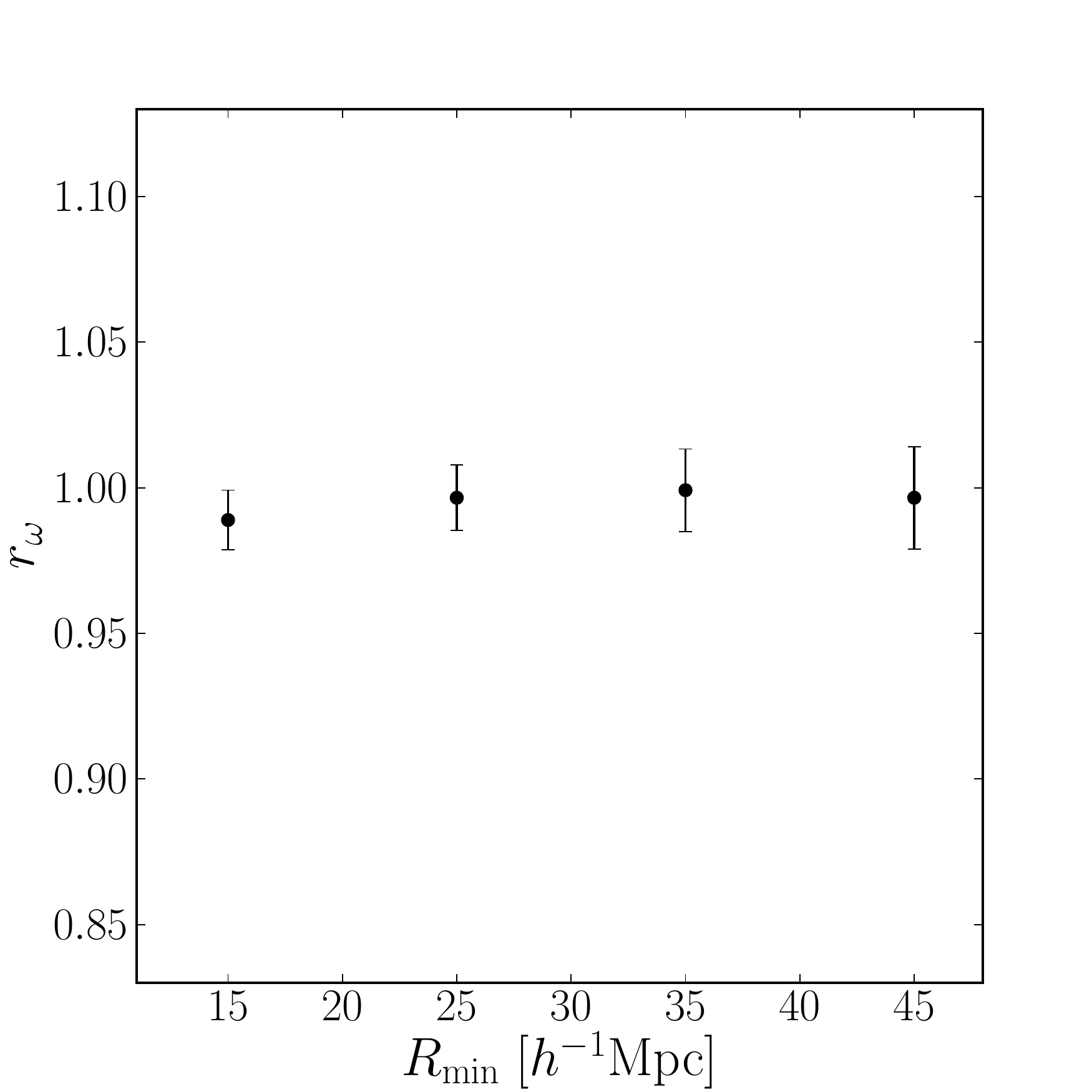}
\end{center}
\caption{The fitted values of $r_{\xi}$ and $r_{\omega}$ as a function of $R_\mathrm{min}$, the minimum radius used in the fit, for one run. The results using the $\xi$ statistic are more strongly dependent on the fitting range than those using $\omega$.}\label{f:fit_var}
\end{figure*}

However, our results averaging over scales are driven primarily by the smaller scale values, as these are better measured in $\xi$ and $\omega$. Accordingly, we test our results by varying the minimum radius and refitting $r_{\xi}$ and $r_{\omega}$ for the same run as depicted in Fig.~\ref{f:xi_r}. The resulting values of $r_{\xi}$ and $r_{\omega}$ are plotted as a function of $R_\mathrm{min}$ in Fig.~\ref{f:fit_var}. From this, we see that $r_{\xi}$ has a much stronger dependence on the minimum radius of the fit than $r_{\omega}$, and shows a steep decline as $R_\mathrm{min}$ increases, which may be due to a systematic at large scales, while $r_{\omega}$ remains roughly constant. This suggests that $\omega$, which demonstrates less variation, is a promising statistic for this type of analysis.
 
For both statistics, using a minimum fitting radius that is $< 20\;h^{-1}\mathrm{Mpc}$ shows results that deviate from what would be obtained using a slightly larger minimum radius. For $\xi$, $r_{\xi}$ is larger, while for $\omega$, $r_{\omega}$ is smaller when $R_\mathrm{min} < 20\;h^{-1}\mathrm{Mpc}$. Accordingly, to avoid results that are driven by these small scales, we recommend using a fitting range $R \gtrsim 20\;h^{-1}\mathrm{Mpc}$, and the results that we quote use that range.

\section{Discussion}
Our results align well with previous studies using correlation function analyses. With data from SDSS DR7, \citet{zehavi11} found that $r_{\xi}$ is consistent with 1 on the largest of scales that they consider (roughly $10 < R < 40\; h^{-1}\mathrm{Mpc}$). This agrees with the results of \citet{wang07}, whose analysis of SDSS DR4 established similar results for $R\sim10\;h^{-1}\mathrm{Mpc}$. Our measured value of $r_{\omega}$ aligns well with these results.

A complementary approach was pursued by \citet{wild05} and \citet{swanson08} in the form of the counts-in-cells method. \citet{wild05} used data from the 2dF Galaxy Redshift Survey to investigate $r_{\xi}$ on scales of $7 < R < 32\; h^{-1}\mathrm{Mpc}$. Their value of $r_{\xi}$ increases from roughly $r_{\xi}\approx0.87$ at $R=7\;h^{-1}\mathrm{Mpc}$ to $r_{\xi}=0.97$ at $R \approx 30\;h^{-1}\mathrm{Mpc}$. \citet{swanson08} relied on SDSS DR5 to obtain similar values for $R\gtrsim25\;h^{-1}\mathrm{Mpc}$, concluding that $r_{\xi}\approx0.97$ as well on such large scales. These results are in good agreement with our values for $r_{\xi}$, but are somewhat lower than what we measure for $r_{\omega}$. Additionally, \citet{blanton00} 
used this method to find that $r_{\xi}\sim0.87-0.95$ for samples of early and late-type galaxies in the Las Campanas Redshift Survey on $15\;h^{-1}\mathrm{Mpc}$ scales, which is consistent with our lower bound from the $\xi$ statistic. 

It is worth noting that in these works as well as our own the correlation coefficient is compared to $r_{\xi}=1$ to determine the level of stochasticity, but in a few bins, our values of $r_{\xi}$ exceed 1. To examine why this is so, we recall the \citet{dekel99} formalism of Equation~(\ref{e:dekeleq}), which describes the red and blue galaxy distributions as:
\begin{align}
\frac{\vec{\delta_r}}{b_r} &= \vec{\delta}+\vec{\epsilon_r} \equiv \vec{\alpha},\\
\frac{\vec{\delta_b}}{b_b} &= \vec{\delta}+\vec{\epsilon_b} \equiv \vec{\alpha}+\vec{\beta}.
\end{align}
Defining a correlation matrix $M$, we may then compute the correlation coefficient as:
\begin{align}
r_{\xi}^2 &= \frac{\langle \vec{\alpha}^TM(\vec{\alpha}+\vec{\beta})\rangle^2}{ \langle \vec{\alpha}^TM\vec{\alpha}\rangle \langle (\vec{\alpha}+\vec{\beta})^TM(\vec{\alpha}+\vec{\beta})\rangle},\\
r_{\xi}^2 &=  \frac{\langle \vec{\alpha}^TM\vec{\alpha}\rangle^2}{\langle \vec{\alpha}^TM\vec{\alpha}\rangle\left[\langle \vec{\alpha}^TM\vec{\alpha}\rangle+\langle \vec{\beta}^TM\vec{\beta}\rangle\right]},
\end{align}
where the cross terms have gone to zero upon taking the expectation value. This means that:
\begin{align}
r_{\xi} = \sqrt{ \frac{\langle \vec{\alpha}^TM\vec{\alpha}\rangle}{\langle \vec{\alpha}^TM\vec{\alpha}\rangle+\langle \vec{\beta}^TM\vec{\beta}\rangle}} \equiv \sqrt{\frac{\xi_{\alpha}}{\xi_{\alpha}+\xi_{\beta}}}.
\end{align}
Accordingly, we have expressed $r_{\xi}$ in terms of the autocorrelation function of $\vec{\beta} = \vec{\epsilon_b}-\vec{\epsilon_r}$. If $\xi_{\beta}=0$, then $r_{\xi}=1$, as expected. However, if $\xi_{\beta}>0$, then $r_{\xi}<1$, but if $\xi_{\beta}$ is negative, then $r_{\xi}$ will in fact exceed 1. 

\section{Comparison with Simulations}
To provide some context for our results, we apply these methods to a catalog produced from a cosmological N-body simulation run with the Abacus code \citep{ferrer,metchnik}, which used $4096^3$ particles in a 3.5 $h^{-1}$Gpc box, yielding particle masses of $5\times10^{10}\:h^{-1}\mathrm{M}_{\odot}$ with a Plummer softening of $105\;h^{-1}\mathrm{kpc}$. Halos at a given redshift were found by the friends-of-friends algorithm \citep{davis85} with a linking length parameter $b=0.2$. We select a full-sky survey of halos, over which we define 145 Voronoi regions, in an annulus corresponding to the range $0.55 < z<0.65$. We extract the real-space centre-of-mass positions of halos that fall into one of two well-separated mass bins, $(0.5-1.0)\times10^{13}\;h^{-1}\mathrm{M_{\odot}}$ and $(0.5-1.0)\times10^{14}\;h^{-1}\mathrm{M_{\odot}}$. These mass ranges do not correspond to the masses of the CMASS galaxies; rather, the lower mass bin is selected so as to have at least 100 particles in the halo, while the other is chosen to be significantly more massive. This split yields 2,183,762 low mass and 113,740 high mass halos. 

\begin{figure}
\begin{center}
\includegraphics[scale=0.5]{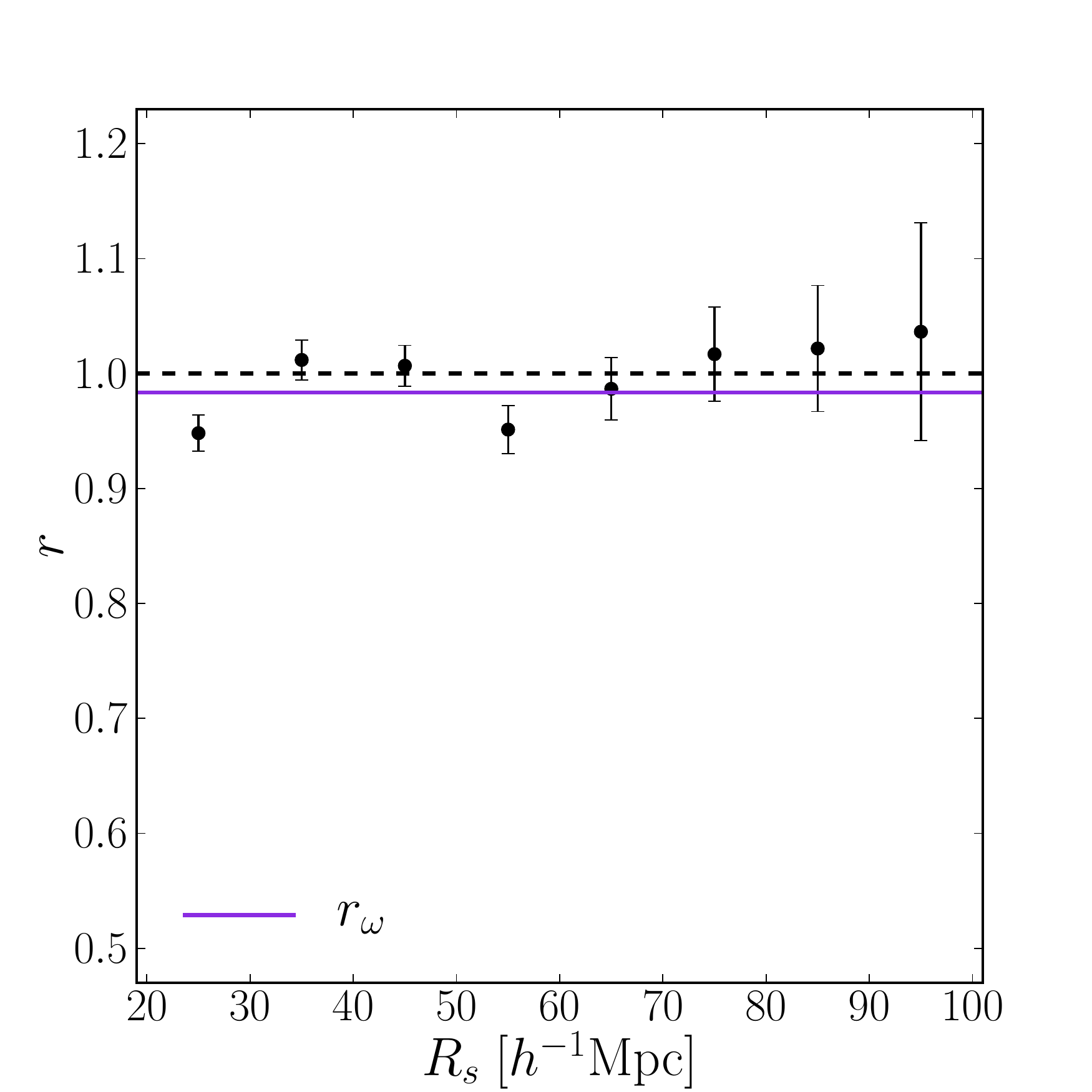}
\end{center}
\caption{The coefficient $r_{\omega}$ for the simulated catalog from the best fit line (solid) shown. A dashed line is drawn at $r=1$ for comparison.}\label{f:xu_r_sim}
\end{figure}

\begin{figure}
\begin{center}
\includegraphics[scale=0.48]{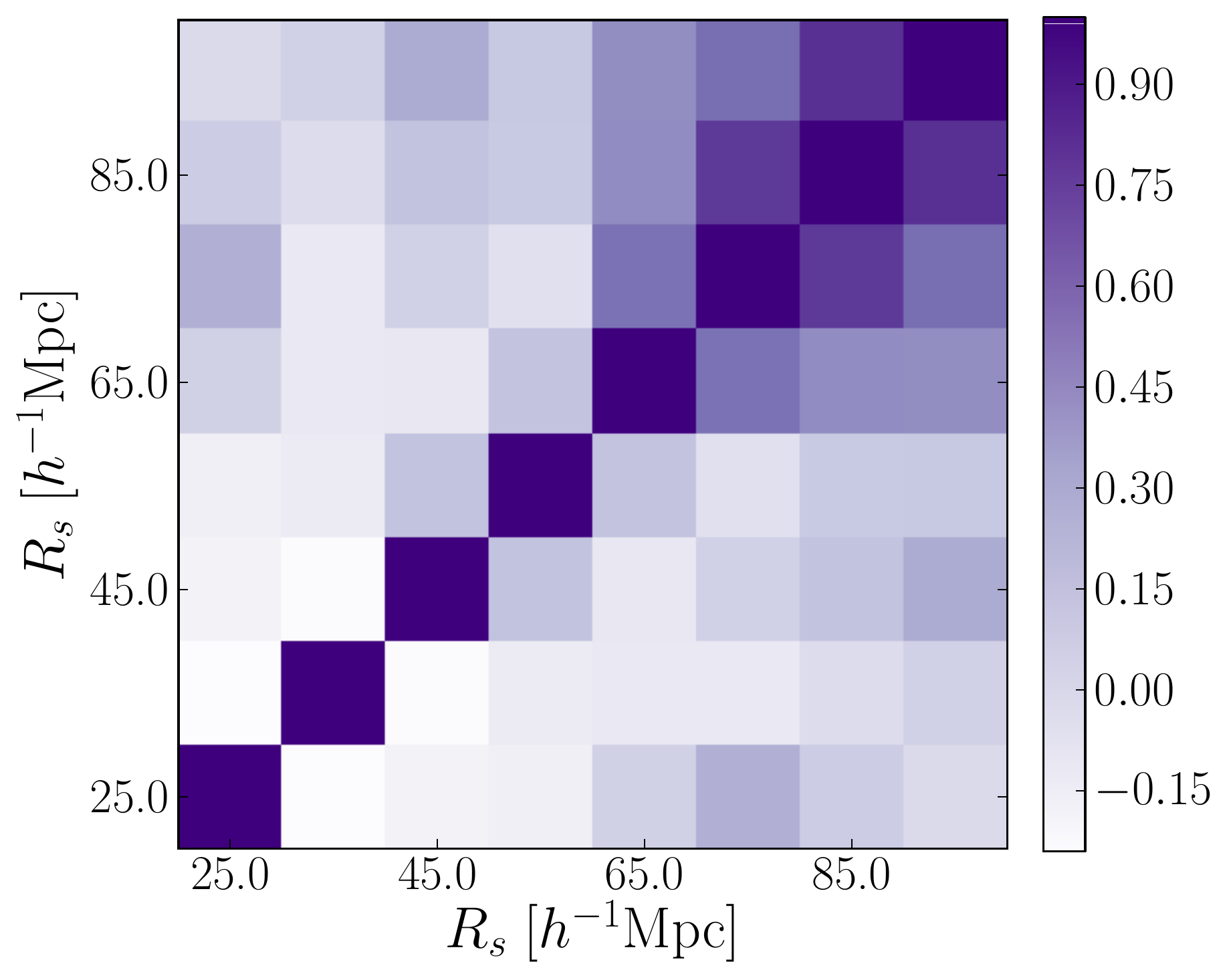}
\end{center}
\caption{The reduced covariance matrix for the correlation coefficient coefficient, $r_{\omega}$, measured from the simulated catalogs, which is used for calculating the $\chi^2$ for the fit shown in Fig.~\ref{f:xu_r_sim}.}\label{f:xu_cov_sim}
\end{figure}

We then follow the same procedure as outlined in Section~\ref{s:method} to measure the coefficient $r_{\omega}$, although we only use one run for the simulated catalog, and only $10\times$ randoms. Fig.~\ref{f:xu_r_sim} shows the variation of $r_{\omega}$ as a function of $R_s$, as well as the best fit horizontal line to the points, and Fig.~\ref{f:xu_cov_sim} shows the reduced covariance matrix used for the fitting. We find a best fit value of $r_{\omega}=0.984\pm0.006$. This measured value is slightly lower than what we found in the data using this same statistic, although the ranges of values are consistent with each other, and with the value of $r_{\xi}$ that was calculated for DR12. However, as noted above, the samples that we use here are not directly comparable to the red/blue split used in the data as these halos are more massive than the CMASS galaxies. Instead, what the simulation results indicate is that a halo model does involve some stochasticity, and provide a lower bound on $r$, which would likely be higher for halo masses more closely attuned to CMASS.

\section{Summary and Conclusions}
Using the CMASS sample of galaxies from SDSS DR12, we have measured the clustering of red and blue galaxies using correlation functions and the $\omega_0$ statistic of \citet{xu10}. From these functions, we constructed the stochasticity parameter $r$ and analysed its value over intermediate scales, which we take as roughly $20 \lesssim R \lesssim 100\;h^{-1}\mathrm{Mpc}$. Basing our results primarily on the $\omega$ statistic, we found that $r>0.974$, indicating low levels of stochasticity on these scales. Taking into account the results using the standard correlation function statistics, this bound decreases slightly to about $r > 0.95$. 

Our results are in good agreement with the results of previous correlation function studies, including those of \citet{wang07} and \citet{zehavi11}. We also compared these values to the value of $r_{\omega}$ obtained from a simulated halo catalog. We found that the simulated catalog prefers a slightly lower value of $r_{\omega}$ than does the data; nevertheless, the values are consistent with each other as well as with the results of using the $\xi$ statistic on the data. However, due to the high mass ranges of the simulated halos, the value of $r$ from this simulation may be lower than what we would obtain from halos with masses more closely corresponding to those of CMASS galaxies.

We have additionally established the utility of the $\omega_0$ statistic in analyzing the stochasticity. As the results of Section \ref{s:results} indicate, the traditional correlation function statistics yield noisy estimates of $r$ on large scales. In addition to being more computationally efficient, the $\omega_0$ statistic is less susceptible to these fluctuations, making it a promising tool for future investigations, possibly at larger scales.

\section*{Acknowledgments}
We would like to thank Ashley Ross, Michael Strauss, and Idit Zehavi for helpful comments on a draft of this paper. We are also grateful to Douglas Ferrer, who produced the Abacus catalog that was used in this paper. 

This work was supported by the National Science Foundation Graduate Research Fellowship under Grant No. DGE-1144152 and by U.S. Department of Energy Grant No. DE-SC0013718. The analysis in this paper made use of tools from NumPy \citep{numpy11}, SciPy \citep{oliphant07}, and Matplotlib \citep{hunter07}, as well as from ROOT \citep[][see also http://root.cern.ch/]{root}.

This paper relies on data from SDSS-III. Funding for SDSS-III has been provided by the Alfred P. Sloan Foundation, the Participating Institutions, the National Science Foundation, and the U.S. Department of Energy Office of Science. The SDSS-III web site is http://www.sdss3.org/.

SDSS-III is managed by the Astrophysical Research Consortium for the Participating Institutions of the SDSS-III Collaboration including the University of Arizona, the Brazilian Participation Group, Brookhaven National Laboratory, Carnegie Mellon University, University of Florida, the French Participation Group, the German Participation Group, Harvard University, the Instituto de Astrofisica de Canarias, the Michigan State/Notre Dame/JINA Participation Group, Johns Hopkins University, Lawrence Berkeley National Laboratory, Max Planck Institute for Astrophysics, Max Planck Institute for Extraterrestrial Physics, New Mexico State University, New York University, Ohio State University, Pennsylvania State University, University of Portsmouth, Princeton University, the Spanish Participation Group, University of Tokyo, University of Utah, Vanderbilt University, University of Virginia, University of Washington, and Yale University.

\bsp	
\label{lastpage}


\begin{thebibliography}{}
\bibitem[Ade et al. (2015)]{planck_cosmo}
Ade P. A. R. et al. (Planck Collaboration), 2015, arXiv: 1502.01589
\bibitem[Alam et al.(2015)]{sdssdr12}
Alam S. et al., 2015, ApJS, 219, 12
\bibitem[Anderson et al.(2012)]{anderson12}
Anderson L. et al., 2012, MNRAS, 427, 3435
\bibitem[Blanton(2000)]{blanton00}
Blanton M., 2000, ApJ, 544, 63
\bibitem[Bolton et al.(2012)]{bolton12}
Bolton A. S. et al., 2012, AJ, 144, 144
\bibitem[Brun \& Rademakers(1997)]{root}
Brun R., Rademakers F., 1997, Nucl. Inst. \& Meth. in Phys. Res. A, 389, 81 
\bibitem[Coil et al.(2008)]{coil08}
Coil A. L. et al., 2008, ApJ, 672, 153
\bibitem[Coles(1993)]{coles93}
Coles P., 1993, MNRAS, 262, 1065
\bibitem[Croton et al.(2007)]{croton07}
Croton D. J., Norberg P., Gazta\~{n}aga E., Baugh C. M., 2007, MNRAS, 379, 1562
\bibitem[Davis et al.(1985)]{davis85}
Davis M., Efstathiou G., Frenk C. S., White S. D. M., 1985, ApJ, 292, 371
\bibitem[Dawson et al.(2013)]{dawson13}
Dawson K. S. et al., 2013, AJ, 145, 10
\bibitem[Dekel \& Lahav(1999)]{dekel99}
Dekel A., Lahav O., 1999, ApJ, 520, 24
\bibitem[Doi et al.(2010)]{doi10}
Doi M. et al., 2010, AJ, 139, 1628
\bibitem[Eisenstein et al.(2011)]{eisenstein11}
Eisenstein D. J. et al., 2011, AJ, 142, 72
\bibitem[Ferrer et al.(in prep)]{ferrer}
Ferrer D. et al. in prep.
\bibitem[Fukugita et al.(1996)]{fukugita96}
Fukugita M. et al., 1996, AJ, 111, 1748
\bibitem[Gunn et al.(1998)]{gunn98}
Gunn J. E. et al., 1998, AJ, 116, 3040
\bibitem[Gunn et al.(2006)]{gunn06}
---, 2006, AJ, 131, 2332
\bibitem[Guo et al.(2013)]{guo13}
Guo H. et al., 2013, ApJ, 767, 122
\bibitem[Hamaus et al.(2010)]{hamaus10}
Hamaus N. et al., 2010, PhRvD, 82, 043515
\bibitem[Hunter(2007)]{hunter07}
Hunter J. D., 2007, Computing in Science \& Engineering, 9, 90
\bibitem[Landy \& Szalay(1993)]{landy93}
Landy S. D., Szalay, A. S., 1993, ApJ, 412, 64
\bibitem[Levi et al.(2013)]{levi13}
Levi M. et al. (DESI Collaboration), 2013, arXiv: 1308:0847
\bibitem[Lupton et al.(2001)]{lupton01}
Lupton R., Gunn J.E., Ivezic Z., Knapp G., Kent S., 2001, Astronomical Data Analysis Software and Systems X, v.238, 269
\bibitem[Masters et al.(2011)]{masters11}
Masters K. L. et al., 2011, MNRAS, 418, 1055
\bibitem[Metchnik \& Pinto(in prep.)]{metchnik}
Metchnik M., Pinto P. in prep.
\bibitem[Oliphant(2007)]{oliphant07}
Oliphant T. E., 2007, Computing in Science \& Engineering, 9, 10
\bibitem[Padmanabhan et al.(2007)]{padmanabhan07}
Padmanabhan N., White M., Eisenstein D. J., 2007, MNRAS, 376, 1702
\bibitem[Padmanabhan et al.(2008)]{padmanabhan08}
Padmanabhan N. et al., 2008, ApJ, 674, 1217
\bibitem[Pier et al.(2003)]{pier03}
Pier J. R. et al., 2003, AJ, 125, 1559
\bibitem[Reid et al.(2015)]{reid15}
Reid B. et al., 2015, arXiv:1509.06529
\bibitem[Ross et al.(2014)]{ross14}
Ross A. J. et al., 2014, MNRAS, 437, 1109
\bibitem[Scherrer \& Weinberg(1998)]{scherrer98}
Scherrer R. J., Weinberg D. H., 1998, ApJ, 504, 607
\bibitem[Seljak \& Warren(2004)]{seljak04}
Seljak U., Warren M. S., 2004, MNRAS, 355,129
\bibitem[Skibba et al.(2014)]{skibba14}
Skibba R. A. et al., 2014, ApJ, 784, 128
\bibitem[Smee et al.(2013)]{smee13}
Smee S. A. et al., 2013, AJ, 146, 32
\bibitem[Smith et al.(2002)]{smith02}
Smith J. A. et al., 2002, AJ, 123, 2121
\bibitem[Suzuki et al.(2012)]{suzuki12}
Suzuki N. et al., 2012, ApJ, 746, 85S
\bibitem[Swanson et al.(2008)]{swanson08}
Swanson M. E. C., Tegmark M., Blanton M., Zehavi I., 2008, MNRAS, 385, 1635
\bibitem[Tegmark \& Bromley (1999)]{tegmark99}
Tegmark M. \& Bromley B. C., 1999, ApJL, 518, L69
\bibitem[van der Walt, Colbert, \& Varoquaux(2011)]{numpy11}
van der Walt S., Colbert S. C., Varoquaux G., 2011, Computing in Science \& Engineering, 13, 22
\bibitem[Wang et al.(2007)]{wang07}
Wang Y., Yang X., Mo H. J., van den Bosch F. C., 2007, ApJ, 664, 608
\bibitem[Weaver et al.(2015)]{weaver15}
Weaver B. et al., 2015, PASP, 127, 397
\bibitem[Wild et al.(2005)]{wild05}
Wild V. et al., 2005, MNRAS, 356, 247
\bibitem[Xu et al.(2010)]{xu10}
Xu X. et al., 2010, ApJ, 718, 1224
\bibitem[York et al.(2000)]{york00}
York D. G. et al., 2000, AJ, 120, 1579
\bibitem[Zehavi et al.(2005)]{zehavi05}
Zehavi I. et al., 2005, ApJ, 621, 22
\bibitem[Zehavi et al.(2011)]{zehavi11}
Zehavi I. et al., 2011, ApJ, 736, 59
\end{thebibliography}
\end{document}